# An SU(3) Unified Model of Electroweak Interaction Using Generalized Yang- Mills Theory

HUANG Si-Zhao [*], LIANG Xiao and WANG Dian-Fu [†]
School of Science, Dalian Maritime University, Dalian 116026, China

***Abstract*** *Generalized Yang-Mills theory has a covariant derivative which contains both vector and scalar gauge bosons. Based on this theory, we construct an $SU(3)$ unified model of weak and electromagnetic interactions. By using the NJL mechanism, the symmetry breaking can be realized dynamically. The masses of $W^{\pm}$, $Z$ are obtained and interactions between various particles are the same as that of Weinberg-Salam (WS) model. At the same time, $\sin^2 \theta_W = 1/4$ can be given.*

## 1 Introduction

Up to now, many experimental results have proved that Weinberg-Salam (WS) model[1,2] is correct in the current energy range. Despite the great success of WS model, as Weinberg himself pointed out, WS model still has some unsatisfactory points[3]. The WS model describes the weak and electromagnetic interactions in the energy range $\leq 10^2 \, GeV$ with two different coupling constants $g$ and $g'$ for the gauge groups $SU(2)$, and $U(1)$, respectively. Thus there is no real explanation of the different strengths displayed by the two interactions. In addition, on the one hand, the experimentally determined Weinberg angle is approximately equal to $30°$, which cannot be directly obtained by WS model itself. On the other hand, although the $125 \, GeV$ Higgs boson has been discovered in 2012[4,5], there is still no evidence that Higgs particles are basic or compound particles and the number of Higgs particles without theoretical guidance. Therefore, the improvement of WS model is still necessary.

There have been different types of ideas to improve these situations. The most widely accepted one by far has been to use a large group of which $SU(2) \otimes U(1)$ is just a small subgroup. The original work is an old idea proposed by Fairlie[6] and Ne'eman[7], of using supergroup $SU(1/2)$ as the unification group for WS model and putting the Higgs fields in the adjoint along with the vector fields. But it increases the dimensions of space-time, meanwhile, the number of the Higgs bosons increases, which is not expected to be seen in theory. In Ref.[8], the authors have constructed an $SU(3)$ unified model of electroweak interaction, by using a special realizations of $SU(3)$ algebras, the correct quantum numbers of the leptons and the Weinberg angle can be given. However, since $SU(3)$ group has eight generators, there are four more vector gauge fields $V^{\pm}$ and $U^{\pm}$ than WS model, as well as some heavy fermions and scalar particles in the model.

Then, is there a more natural way for us to introduce the Higgs fields to physical theories? Many scholars have taken efforts to solve this problem. In Ref.[9], the authors have constructed a $U(3)$ unified model of electroweak interaction using a generalized Dirac covariant derivative, that contained both vector and pseudo-scalar fields. However, since $U(3)$ group has nine generators, it will have a extra field that does not interact with other particles than WS model. And what's more,

WANG Dian-Fu [†], E-mail: wangdfu@dlmu.edu.cn
HUANG Si-Zhao [*], E-mail: hsz@dlmu.edu.cn

there is no the potential energy term in the model, thus the spontaneous symmetry breaking cannot be applied, the particles in the model cannot obtain masses. Recently some authors have attempted to construct the so-called generalized Yang–Mills theory (GYMT)[10-14], which the generalized Dirac covariant derivative $D$ besides the vector part $A_\mu$, it can also contains a scalar part $\varphi$, a pseudo-scalar part $P$, an axial-vector part $V_\mu$ and a tensor part $T_{\mu\nu}$. In Ref.[12], by using a covariant derivative with both vector and scalar gauge fields, the authors have constructed a generalized Yang–Mills model, which is invariant under local gauge transformations of a Lie group. Since the GYMT does not involve the potential energy term about the scalar fields, it is difficult to realize the Higgs mechanism[15] directly. It is shown, in terms of the NJL mechanism[16], that the gauge symmetry breaking can be realized dynamically.

Based on the GYMT given in Ref.[12], the work of the present paper is to construct an $SU(3)$ gauge-invariant unified model of electroweak interaction and that it naturally assigns the correct quantum numbers to the leptons and Higgs bosons. By using the GYMT, we introduce vector fields and scalar fields as the gauge fields into the model by the requirement of localization gauge invariance. We will show that, in terms of the NJL mechanism, the symmetry breaking can be realized dynamically and the masses of $W^\pm$ and $Z$ particles are obtained. Meanwhile, interactions between various particles are the same as that of WS model.

## 2 Generalized Yang-Mills Theory

The main idea of the GYMT in Ref.[12] is as follows: Corresponding to each generator of the Lie group there is a gauge field, it does not matter whether vector fields or scalar fields. The generalized Dirac covariant derivative $D$ can be constructed by taking each of the $N$ generators and multiplying it by one of its associated gauge fields and summing them together

$$D = \gamma_\mu \partial_\mu - i\gamma_\mu A_\mu + \varphi, \tag{1}$$

where

$$A_\mu = gA_\mu^a T_a, \quad \varphi = g\varphi^b T_b, \tag{2}$$

with the subscript $a$ varies from $1$ to $N_A$, $b$ varies from $N_A+1$ to $N$. Define the transformation for the gauge fields as

$$-i\gamma_\mu A_\mu + \varphi \to U\left(-i\gamma_\mu A_\mu + \varphi\right)U^{-1} - \left(\gamma_\mu \partial_\mu U\right)U^{-1}, \tag{3}$$

from which we can obtain that the covariant derivative must transform as $D \to UDU^{-1}$. When the covariant derivative acts on the matter field $\psi$, its gauge fields will acquire certain charge of each gauge field with respect to $\psi$ with the result

$$D_\psi = \gamma_\mu \partial_\mu - iQ_A \gamma_\mu A_\mu + Q_\varphi \varphi. \tag{4}$$

Based on the standard model, we can conclude that $Q_A = 1$.

The Lagrangian of the GYMT contains only covariant derivatives and matter fields, and that it possesses both gauge and Lorentz invariance:

$$L = -\bar\psi D_\psi \psi + \frac{1}{2g^2}\tilde{Tr}\left(\frac{1}{8}\left(TrD^2\right)^2 - \frac{1}{2}TrD^4\right), \tag{5}$$

In which the trace with the tilde is over the matrices of the Lie group and the one without tilde is over the matrices of the spinorial representation of Lorentz group.

## 3 The SU(3) Unified Model of Electroweak Interaction

In this section, in terms of the above GYMT, we will construct an $SU(3)$ unified model of weak and electromagnetic interactions. By considering an $SU(3)$ gauge invariant GYMT, it will be naturally assign the correct isospin $T_3$ and hypercharge $Y$ quantum numbers to the neutrino and the electron, so long as we place them in $SU(3)$ fundamental representation

$$\psi = \begin{pmatrix} \nu_L \\ e_L \\ e_R \end{pmatrix}. \tag{6}$$

In the Lagrangian (5), the covariant derivative $D$ will be of the form Eq.(1) where $A_\mu = gA_\mu^a T_a$ $(a=1,2,3,8)$ are the vector gauge fields, $\varphi = g\varphi^b T_b$ $(b=4,5,6,7)$ are the scalar gauge fields, and $g$ is the coupling constant, we have

$$D = \gamma_\mu \partial_\mu - i\gamma_\mu g A_\mu^a T_a + g\varphi^b T_b, \tag{7}$$

in which $T_i = (1/2)\lambda_i$ are the generators of $SU(3)$ group in three-dimensional representation, the $\lambda_i$ are $3\times 3$ traceless hermitian matrices, which can be chosen to have the form

$$\lambda_1 = \begin{pmatrix} 0 & 1 & 0 \\ 1 & 0 & 0 \\ 0 & 0 & 0 \end{pmatrix}, \lambda_2 = \begin{pmatrix} 0 & -i & 0 \\ i & 0 & 0 \\ 0 & 0 & 0 \end{pmatrix}, \lambda_3 = \begin{pmatrix} 1 & 0 & 0 \\ 0 & -1 & 0 \\ 0 & 0 & 0 \end{pmatrix}, \lambda_4 = \begin{pmatrix} 0 & 0 & 1 \\ 0 & 0 & 0 \\ 1 & 0 & 0 \end{pmatrix},$$

$$\lambda_5 = \begin{pmatrix} 0 & 0 & -i \\ 0 & 0 & 0 \\ i & 0 & 0 \end{pmatrix}, \lambda_6 = \begin{pmatrix} 0 & 0 & 0 \\ 0 & 0 & 1 \\ 0 & 1 & 0 \end{pmatrix}, \lambda_7 = \begin{pmatrix} 0 & 0 & 0 \\ 0 & 0 & -i \\ 0 & i & 0 \end{pmatrix}, \lambda_8 = \frac{1}{\sqrt{3}}\begin{pmatrix} -1 & 0 & 0 \\ 0 & -1 & 0 \\ 0 & 0 & 2 \end{pmatrix}. \tag{8}$$

One can see that the matrices from $\lambda_1$ to $\lambda_7$ are still the usual Gell-Mann matrices. Meanwhile the last one $\lambda_8$ takes the minus sign. Correspondingly, the components containing indices 8 in the structure constants $f_{ijk}$ and $d_{ijk}$ of the $SU(3)$ group will take the minus sign too.

Next, we will give the specific form of the covariant derivative $D_\psi$ in Eq.(5). Following from Ref.[8], in order to obtain the correct hypercharge of $e_R$, we choose one of the four realizations of the $SU(3)$ algebra. Then the above eight generators $T_i$ can be divided into two groups $T_n$ $(n=2,4,6)$ and $T_s$ $(s=1,3,5,7,8)$ respectively. Define $T_i^{(5)}$ $(i=1,\ldots,8)$ as

$$T_n^{(5)} = T_n, \quad T_s^{(5)} = T_s \gamma_5. \tag{9}$$

It can be easily proved that $T_i^{(5)}$ satisfy the same commutation rules as $T_i$

$$\left[T_i^{(5)}, T_j^{(5)}\right] = if_{ijk} T_k^{(5)}. \tag{10}$$

Here, $T_i^{(5)}$ is the one of the four realizations of $SU(3)$ algebra.

Following the above discussion we can give the covariant derivative $D_\psi$ as

$$D_\psi = \gamma_\mu \partial_\mu - i\gamma_\mu A_\mu^{(5)} + Q_\varphi \varphi^{(5)}, \tag{11}$$

where $A_\mu^{(5)} = gA_\mu^a T_a^{(5)}$ $(a=1,2,3,8)$, $\varphi^{(5)} = g\varphi^b T_b^{(5)}$ $(b=4,5,6,7)$. By using $\gamma_5 L = L$, and $\gamma_5 R = -R$, we can obtain

$$i\bar{\psi}\gamma_\mu A_\mu^{(5)}\psi = ig\bar{\psi}\gamma_\mu \left[\gamma_5\left(A_\mu^1 T_1 + A_\mu^3 T_3 + A_\mu^8 T_8\right) + A_\mu^2 T_2\right]\psi$$

$$= ig\bar{\psi}\gamma_\mu \left[A_\mu^1 T_1 + A_\mu^2 T_2 + A_\mu^3 T_3 + A_\mu^8 T_8'\right]\psi, \tag{12}$$

in which $T_8' = 1/2\sqrt{3}\, diag(-1,-1,-2)$. This means that the hypercharges of $\nu_L$, $e_L$ and $e_R$ are

$-1,-1,-2$, respectively.

Substituting Eq.(12), Eq.(11) and Eq.(7) into Eq.(5), we obtain the Lagrangian

$$L = -\bar{\psi}\gamma_\mu \partial_\mu \psi + i\bar{\psi}\gamma_\mu A_\mu^{(5)} \psi - Q_\varphi \bar{\psi}\varphi^{(5)}\psi - \frac{1}{2g^2}\tilde{Tr}\left(\partial_\mu A_\nu - \partial_\nu A_\mu - i\left[A_\mu, A_\nu\right]\right)^2$$

$$-\frac{1}{g^2}\tilde{Tr}\left(\partial_\mu \varphi - i\{A_\mu, \varphi\}\right)^2. \tag{13}$$

By using the Pauli matrices $\sigma_{a'}$ ( $a'=1,2,3$ ), Eq.(6) and Eq.(12), we can give the expansion of the Lagrangian as

$$L = -\bar{\theta}_L \gamma_\mu \left(\partial_\mu - \frac{1}{2}igA_\mu^{a'}\sigma_{a'} + \frac{1}{2}ig'A_\mu^8\right)\theta_L - \bar{e}_R\gamma_\mu\left(\partial_\mu + ig'A_\mu^8\right)e_R$$

$$-\frac{G_F}{\sqrt{2}}\bar{e}_R\phi^{(5)+}\theta_L - \frac{G_F}{\sqrt{2}}\bar{\theta}_L\phi^{(5)}e_R - \frac{1}{4}F_{\mu\nu}^{a'}F_{\mu\nu}^{a'} - \frac{1}{4}F_{\mu\nu}^8 F_{\mu\nu}^8$$

$$-\left|\left(\partial_\mu + \frac{1}{2}igA_\mu^{a'}\sigma_{a'} + \frac{1}{2}ig'A_\mu^8\right)\phi\right|^2, \tag{14}$$

where, $F_{\mu\nu}^{a'} = \partial_\mu A_\nu^{a'} - \partial_\nu A_\mu^{a'} + gf_{a'b'c'}A_\mu^{b'}A_\nu^{c'}$ ( $a',b',c'=1,2,3$ ), $F_{\mu\nu}^8 = \partial_\mu A_\nu^8 - \partial_\nu A_\mu^8$, $\theta_L = (\upsilon_L, e_L)^T$, $\phi = (\varphi^4 - i\varphi^5, \varphi^6 - i\varphi^7)^T/\sqrt{2}$, $\phi^{(5)} = (\varphi^4 - i\gamma_5\varphi^5, \varphi^6 - i\gamma_5\varphi^7)^T/\sqrt{2}$, $G_F = gQ_\varphi$ and $g' = g/\sqrt{3}$. By considering $g' = g/\sqrt{3}$, one can conclude that $\sin^2\theta_W = 1/4$ easily. Seeing from Eq.(14), one can obtain the correct hypercharge $Y=1$ of the scalar fields (Higgs particles) as in WS model.

## 4 Dynamical Breaking of SU(3) Gauge Symmetry

As is known to us, Eq.(14) is almost the Lagrangian of WS model, except that no the Higgs potential $V(\varphi)$, which means that the spontaneous symmetry breaking mechanism cannot be utilized directly. In this section, we will show that by using the NJL mechanism, The $SU(3)$ gauge symmetry breaking can be realized dynamically.

Substituting Eq.(13) into Euler equation, one can obtain the equations of motion for the fermion fields $\psi$, the scalar gauge fields $\varphi^b$, and the vector gauge fields $A_\mu^{a'}$

$$\gamma_\mu\left(\partial_\mu - igA_\mu^a T_a^{(5)}\right)\psi + G_F \varphi^b T_b^{(5)}\psi = 0, \tag{15}$$

$$\left(\partial_\mu^2 - g^2 d A_\mu^a A_\mu^a\right)\varphi^b - G_F \bar{\psi}T_b^{(5)}\psi = 0, \tag{16}$$

$$\left(\partial_\mu F_{\mu\nu}^{a'} + gf_{a'b'c'}A_\mu^{b'}F_{\mu\nu}^{c'}\right) + g^2 d\left(\varphi^b\right)^2 A_\nu^{a'} - ig\bar{\psi}\gamma_\nu T_a^{(5)}\psi = 0. \tag{17}$$

With $d = d_{abc}d_{abc}$ ( $a=1,2,3,8$, $b,c=4,5,6,7$ ). Multiplying $A_\nu^{a'}$ on both sides of Eq.(17), we obtain

$$\left[\left(\partial_\mu F_{\mu\nu}^{a'} + gf_{a'b'c'}A_\mu^{b'}F_{\mu\nu}^{c'}\right) + g^2 d\left(\varphi^b\right)^2 A_\nu^{a'} - ig\bar{\psi}\gamma_\nu T_a^{(5)}\psi\right]A_\nu^{a'} = 0, \tag{18}$$

after taking the vacuum expectation value of Eq.(18), to the lowest-order approximation in $\hbar$, we obtain a simple formula

$$f\left\langle A_\mu^{a'}A_\mu^{a'}\right\rangle = d\left\langle\left(\varphi^b\right)^2\right\rangle, \tag{19}$$

where $f = f_{a'b'c'}f_{c'b'a'}$ ( $a',b',c'$ from 1 to 3 ). We can see that in the ground state, Eq.(19) gives an important relationship between the vector gauge fields and the scalar gauge fields. One can denote the vacuum expectation of the scalar fields as

$$\left\langle\varphi^b\right\rangle = \left\langle\varphi^6\right\rangle = \upsilon \neq 0, \tag{20}$$

which means

$$\langle\phi\rangle = \langle\phi^{(5)}\rangle = \frac{1}{\sqrt{2}}\begin{pmatrix} 0 \\ \upsilon \end{pmatrix}. \tag{21}$$

Substitute Eq.(21) into Eq.(14), we can give the masses of the neutrino and the electron

$$m_e = \frac{G_F \upsilon}{2}, \quad m_\nu = 0. \tag{22}$$

Let us now take the vacuum expectation value of Eq.(16). To the lowest-order approximation in $\hbar$, by using Eq.(19), the self-consistency equation can be given as

$$g^2 d^2 f^{-1} \langle\varphi^6\rangle^3 = -\frac{1}{2} G_F \langle\bar{e}e\rangle. \tag{23}$$

In Eq.(23), with an invariant momentum cut-off at $p^2 = \Lambda^2$ in the momentum integral, $\langle\bar{e}e\rangle$ will be finite value as

$$\begin{aligned}\langle\bar{e}e\rangle &= -TrS_F(0) = 2G_F \langle\varphi^6\rangle \int \frac{d^4 p}{(2\pi)^4} \frac{i}{p^2 + G_F^2 \langle\varphi^6\rangle^2 / 4} \\ &= -\frac{G_F \langle\varphi^6\rangle}{4\pi^2}\left[\Lambda^2 - \frac{G_F^2 \langle\varphi^6\rangle^2}{4} \ln\left(\frac{4\Lambda^2}{G_F^2 \langle\varphi^6\rangle^2} + 1\right)\right].\end{aligned} \tag{24}$$

Substituting Eq.(24) into Eq.(23), we have

$$\langle\varphi^6\rangle^2 = \frac{G_F^2 f}{8\pi^2 d^2 g^2}\left[\Lambda^2 - \frac{G_F^2 \langle\varphi^6\rangle^2}{4} \ln\left(\frac{4\Lambda^2}{G_F^2 \langle\varphi^6\rangle^2} + 1\right)\right]. \tag{25}$$

From Eq.(25), one can finally obtain the non-vanishing vacuum expectation value $\langle\varphi^6\rangle$ of the scalar fields, which is determined by the self-energy of the fermion fields. And then the $SU(3)$ gauge symmetry is broken down dynamically.

Substituting the definition

$$W_\mu^\pm = \frac{1}{\sqrt{2}}(A_\mu^1 \mp i A_\mu^2), \quad Z_\mu = \frac{1}{2}(-\sqrt{3} A_\mu^3 + A_\mu^8), \quad B_\mu = \frac{1}{2}(A_\mu^3 + \sqrt{3} A_\mu^8), \tag{26}$$

and Eq.(21) into Eq.(14), we can obtain the masses of the vector gauge particles

$$m_W^2 = \frac{1}{4} g^2 \upsilon^2, \quad m_Z^2 = \frac{4}{3} m_W^2, \quad m_B^2 = 0. \tag{27}$$

This result is exactly the same as that of the standard model.

## 5 Summary and Remarks

In this paper, based on the generalized Yang-Mills theory, we construct an $SU(3)$ unified model of weak and electromagnetic interactions. By using the NJL mechanism, the $SU(3)$ gauge symmetry breaking can be realized dynamically, although there is no the Higgs potential $V(\varphi)$ in the GYMT. The masses of $W^\pm$ and $Z$ particles are obtained, interactions and quantum numbers of various particles are the same as that of WS model. Compared to WS model, the present model has several advantages. Firstly, since the present model is based on $SU(3)$ gauge group, there is only one coupling constant $g$, and $\sin^2\theta_W = 1/4$ can be obtained directly. Secondly, in the present model the scalar fields (Higgs fields) are considered to be as gauge fields, then the introduction of the scalar fields becomes natural, and the numbers of the scalar fields can becomes certain too.

At the end of this paper, we wish to point out that in terms of Eq.(16) and Eq.(19) we should be able to calculate the masses of the Higgs particles approximately. We will discuss this problem in our forthcoming works.